\RequirePackage{fix-cm}
\documentclass{svjour3}                     
\smartqed  
\usepackage{graphicx}
\usepackage{mathptmx}      
\usepackage{latexsym}
\usepackage{amssymb}
\usepackage{amsmath}
\usepackage{graphicx}
\usepackage{epstopdf}
 \journalname{Mathematical Geoscience}
\begin{document}

\title{Periodic void formation in chevron folds
}


\author{T. J. Dodwell         \and
        G. W. Hunt
}


\institute{T. J. Dodwell \at
              Department of Mechanical Engineering, University of Bath, BA2 7AY \\
              \email{tjd20@bath.ac.uk}           
           \and
           G. W. Hunt \at
              Department of Mechanical Engineering, University of Bath, BA2 7AY \\
              \email{ensgwh@bath.ac.uk}           
}

\date{Received: date / Accepted: date}

\maketitle

\begin{abstract}
An energy-based model is developed to describe the periodic formation of \emph{voids}/\emph{saddle reefs} in  hinge zones of chevron folds. Such patterns have been observed in a series of experiments on layers of paper, as well as in the field. A simplified hinge region in a stack of elastic layers, with straight limbs connected by convex segments, is constructed so that a void forms every $m$ layers and repeats periodically. Energy contributions include strain energy of bending, and work done both against a confining overburden pressure and an axial compressive load. The resulting total potential energy functional for the system is minimized subject to the constraint of non-interpenetration of layers, leading to representation as a nonlinear second-order free boundary problem. Numerical solutions demonstrate that there can exist a minimum-energy $m$-periodic solution with $m \ne 1$. Good agreement is found with experiments on layers of paper.
\keywords{folding \and delamination \and energy methods \and saddle reef formation}
\end{abstract}

\section{Introduction}\label{sec:Introduction}

\begin{figure}
\centering
\includegraphics[height=1.5in]{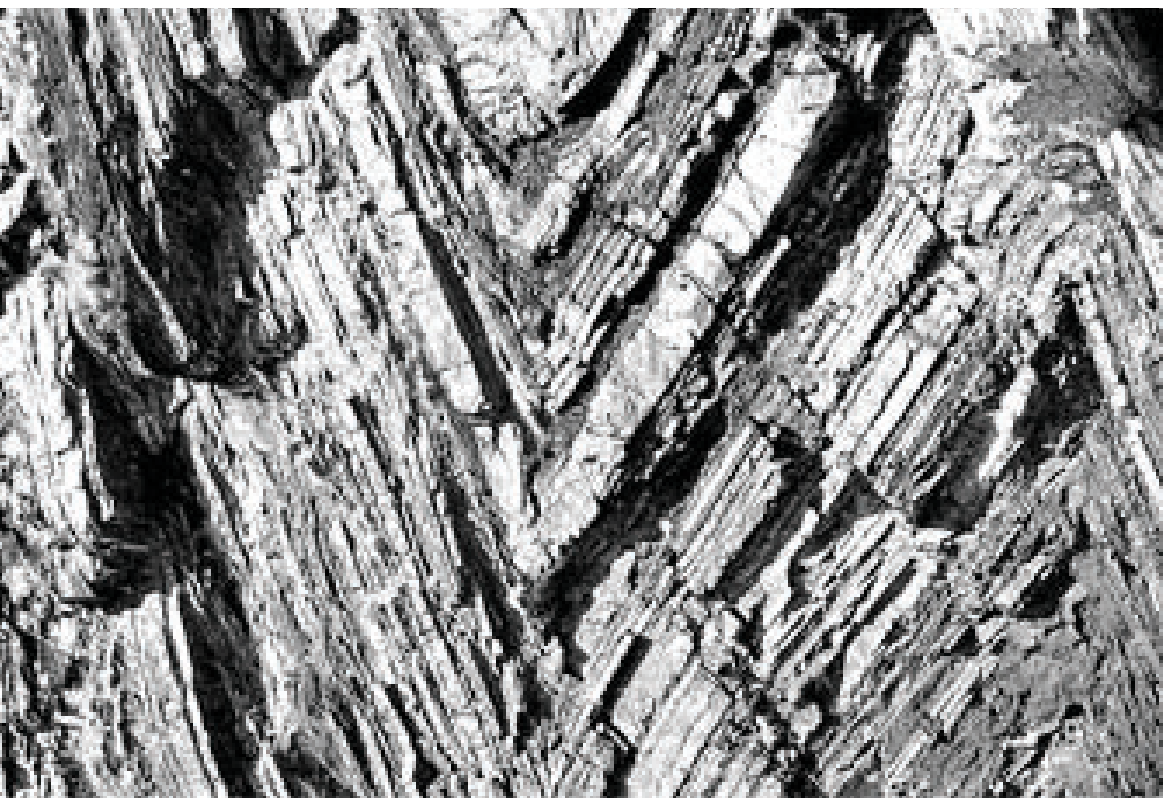}\quad\includegraphics[height=1.5in]{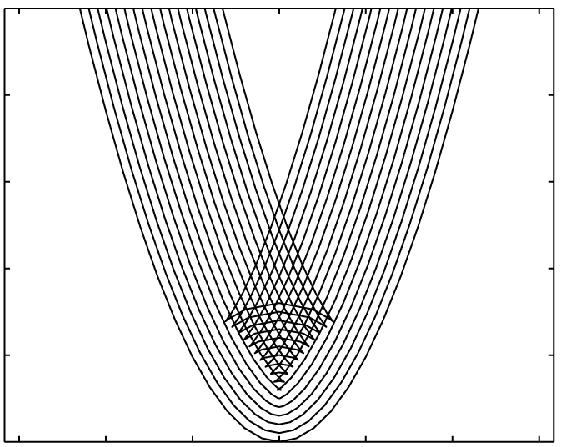}
\caption{(Left) A multilayered folding pattern: chevron folds in thinly bedded shales and sandstones in Millook Haven, Cornwall UK. (Right) Propagation of a parabolic wavefront, demonstrating the physically unrealisable swallowtail catastrophe if normal thickness is constrained to remain constant.}
\label{fig:rocks}
\end{figure}

Multilayered structures found in the Earth's crust can arise from sequences of sedimentary deposits, subsequently held together by the weight of additional overlying layers. The bending and buckling of such layers during tectonic collision will depend on an intricate mix of  complex mechanical and material properties, but also on thoroughly nonlinear constraints coming from the layers being obliged to fit together. These geometric constraints generate a number of distinctive forms of folding not seen in single-layered systems. An example of multilayered folding from Millook Haven in Cornwall, UK is shown in Fig.~\ref{fig:rocks} (left).

\smallskip
If multiple layers of rock are constrained such that they remain in contact during folding, sharp hinges are likely to develop. To see this, we consider first a single layer in the shape of a parabola, Fig.~\ref{fig:rocks}, and then place further layers of constant thickness on top. The addition of new layers means the curvature of each tightens in turn until it becomes infinite, marking the presence of a \emph{swallowtail singularity} (Boon et al. 2007), beyond which layers would need to interpenetrate. Layered systems can accommodate for this singularity by adopting straight limbs punctuated by sharp hinges, as seen in the chevron folds of Fig. \ref{fig:rocks}. But folding into a sharp hinge can be  energetically very expensive, and is often approximated by high rather than infinite curvature regions, separated by distinctively-shaped \emph{voids}, into which material from more highly pressurised regions could leach (Dodwell et al. 2012a). Such veins are termed  \emph{saddle reefs} (Price \& Cosgrove 1990, Ramsay 1974), and are commonly found in vertical succession within larger chevron  folds (Price \& Cosgrove 1990). 

\smallskip
The formation of voids, or similarly regions of reduced pressure, is of particular economic interest, since they frequently form localised areas in which minerals collect. A notable field example is the gold-rich quartz deposits of Bendigo, Australia. Here a sequence of saddle reefs have formed in vertical succession through the hinge region of multiple folds, which have been subsequently filled with quartz and concentrations of gold. The chevron folds at Bendigo have formed directly above a basement that was undergoing prograde metamorphism whilst simultaneously folding. Prograde metamorphism occurs when temperatures and pressures acting on the rock increase; the rocks then undergo a series of chemical reactions, to form minerals more stable under these conditions. As hot fluids rise and subsequently cool closer to the earth surface, large convection currents of fluid, known as Rayleigh-Bernard convection rolls, are generated in the Earth's crust. As the fluid rises and temperatures fall, both minerals and metals become less soluble and can precipitate and be deposited amongst the rock folds. The voids or saddle reefs, multiple accommodation thrusts and the formation of tension fractures in the outer arc of folds, increase the permeability for fluids through the hinge. Over time this process deposits both quartz and gold into the vertical sequence of voids that have formed. 

\smallskip
The mechanics of folding of thinly layered rock systems is a topic of considerable interest in the field of structrual geology. Stemming from the inital geometic observations of multilayered folds most notably by Hall (1815) and Van Hise (1894), various contributions have investigated the mechanics of multilayered folding. In particularly, Biot's contributes on internal buckling instabilites and fold type selection (Biot 1965), Ramberg's concept of `zone contact strain' to described harmonic folding patterns (Ramsay 1964, 1970) and Ramsay's geometric models for Chevron fold development (Ramsay 1974). Motivated by the geometric approach of the latter, the authors developed a simplified energy-based model to describe a single void (Dodwell et al. 2012), in which an elastic layer is forced by a uniform overburden pressure $q$ into a V-shaped singularity. Besides demonstrating various mathematical properties of energy-minimising solutions (convexity, uniqueness and symmetry) it was shown that the horizontal measure of the void size varies as $q^{-1/3}$. Therefore,  with the high overburden pressures found in most geological circumstances (Hobbs et al. 1976,Price and Cosgrove 1990), voids would be expected to be relatively small. The model has been extended to embrace multilayers of finite length and effects of axial loads (Dodwell et al. 2012b), but only in an approximation that assumes constant curvature over the voided regions, and hence contravenes boundary conditions at the points of separation. Here, using the same mathematical toolkit, numerical solutions that ensure that these boundary conditions are correctly satisfied. Classically layers of rock are consider viscous or viscous plastic materials (e.g Biot (1964), Frehner \& Schmalholz (2006), Price \& Cosgrove (1990) and Ramberg (1964,1970)). Purely elastic assumptions are often considered inaccurate for real rock folds; however more complex rheological models are also a vast simplication of their geological counterparts, which are depend on an intricate balance of nonlinear constutive laws, strain histrory, temperature, pressure and chemcial dependence. Such complexities appears almost impregnable to a mathematical analysis. Here the key focus of the paper is to consider the geometric nonlinearities that naturally arise due to the constraints of layers fitting together during their process. The assumptions of the layers are linear elastic allows us to isolate the role that the highly nonlinear geometric constraints of layers fitting together has during multilayered folding processes.

\subsection{Paper experiments}
The sequence of events leading to multilayered geological structures are extremely complex, and it is often  useful to make comparisons against simpler but more readily quantifiable systems. Small scale experiments on layered materials such as paper can provide useful benchmarks, and  earlier comparisons have helped to shed light on some naturally occurring phenomena, such as kink-banding (Wadee et al. 2004) and parallel folding (Budd et al. 2003).

The experiments described in this paper span a number of years (Dodwell et al. 2012, Wadee et al. 2004), but the general setup is much the same. In each case, layers of paper (typically $A5$ or $A6$ sheets) are stacked vertically, with a foam foundation placed on either side. A simple rig made of plywood holds the stack and foundation together, and the rig can be tightened via G-clamps to add some initial pre-stress into the foundation, applying a transverse overburden pressure $q$ per unit length. The rig is placed in an Instron testing machine, and a compressive load $P$ applied at a constant (slow) rate via  a motorized load applicator.

\smallskip
Figure~\ref{fig:voidexperiments} shows four stages during one such experiment, where 750 layers of A6 paper are buckled by a vertical compressive load whilst being constrained laterally by a soft foam foundation.
To aid visualization, a black-edged sheet has been introduced approximately every 25 layers. This experiment demonstrates that voiding can be regular. Although the disturbance of introducing black-edged sheets clearly helps to induce voids, they appear in a periodic sequence every five black sheets, or approximately once every 125 layers. 

In a similar experiment, Fig.~\ref{fig:1periodexample}, with a stiffer surrounding foam, the process of voiding is not as clear. However, the magnified view of the hinge region on the right shows a void developing for every layer. These two examples of periodic voiding patterns in the hinge regions of chevron folds are intriguing, and lead to interesting questions. Can we explain for example, why under some conditions voids form regularly at every layer level, whereas for others they repeat only over a large number of layers?  And quantitatively, using an energy-based argument, can we predict the periodicity of the voiding?

\subsection{Outline of Paper}

\begin{figure}
\centering
\includegraphics[width=5in]{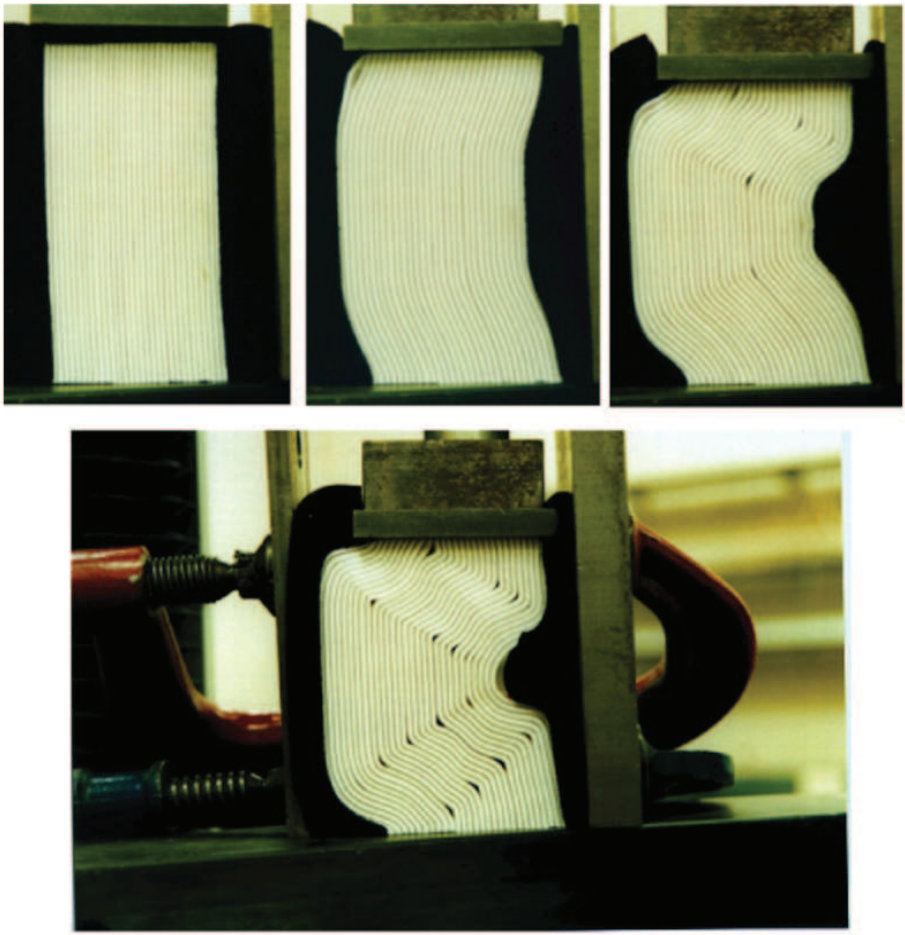}
\caption{A  laboratory experiment of layers of  A6 paper constrained and loaded in-plane. The resulting deformation shows that voids form when curvatures become too high. We particularly note the regular periodic nature of the voids (approximately every 125 layers).    \ }
\label{fig:voidexperiments}
\end{figure}

\begin{figure}
\centering
\includegraphics[width = 5in]{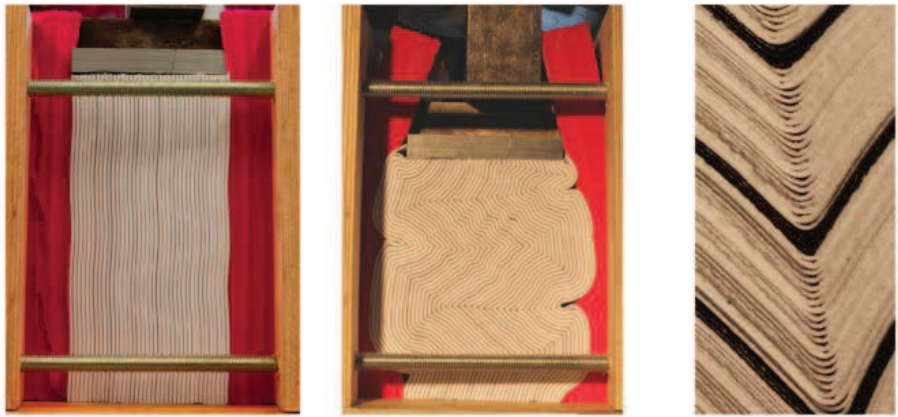}
\caption{A similar experiment to Fig.~\ref{fig:voidexperiments} but with a much stiffer constraining foam. Only on close inspection are the formation of voids visible, where they produce a regular periodic pattern at every layer level. }
\label{fig:1periodexample}
\end{figure}

\begin{figure}
\centering
\includegraphics[width=5in]{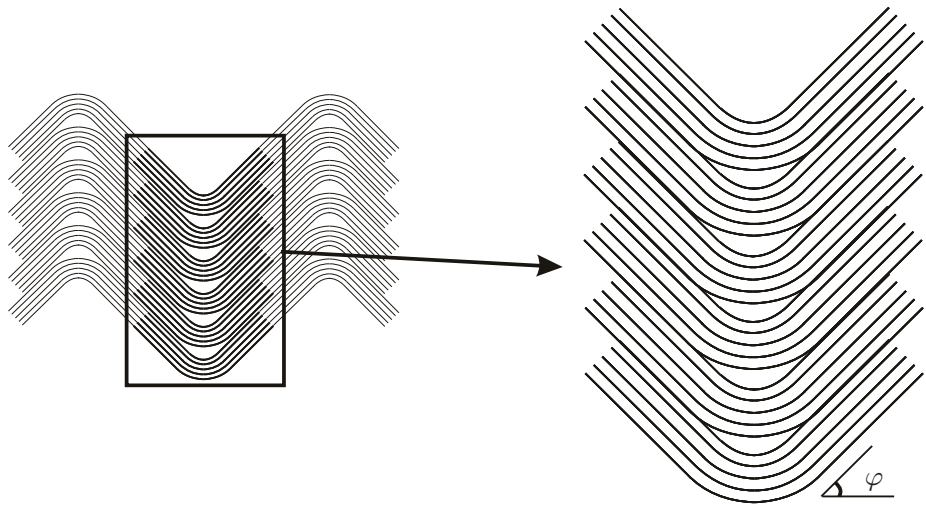}
\caption{Section of a chevron fold. Here the $4$-periodic stack is characterised by straight limbs of angle $\varphi$ to the horizontal, connected by a symmetric convex segment. A single void occurs between each periodic repeat.}
\label{fig:chevronfold}
\end{figure}

To study this periodic behaviour we define an $m$-periodic solution in a vertical stack of layers as one in which a single void forms every $m$ layers, with the same pattern repeating periodically; Fig.~\ref{fig:voidexperiments} shows a 125-periodic solution for example, whereas Fig.~\ref{fig:1periodexample} is a 1-period solution. We present a simplified energy-based model for an $N$-layered section of a large chevron fold comprising identical elastic layers, arranged into an $m$-periodic stack (See Fig.~\ref{fig:chevronfold}). The geometry of the chevron fold is assumed to have straight limbs at angle $\varphi =\tan^{-1}k$, connected by a convex curved segment. A uniform overburden pressure $q$ holds the layers together, while a compressive axial load $P$ acts to shorten each layer.

\smallskip
The resulting profiles are obtained by finding convex symmetric energy minimisers of the potential energy functional for the system (Section~\ref{subsec:TPEandEL}), under the constraint that layers do not interpenetrate. It is assume that layers store energy in bending and against overburden pressure in voiding, while work is done by the axial load $P$. For this initial simplified model, interactions between layers are considered to be under hard frictionless contact, and constrained energy minimization leads to its representation as a second-order free boundary problem.

\smallskip
Section~\ref{subsec:numerics} shows that numerical solutions to this free-boundary problem agree with physical intuition, in that with increase in either overburden pressure $q$ or axial load $P$, the size of the voids in the multilayered stack decreases. Additionally, by finding the energy (\ref{eqn:potential}) of the solutions for various values of $m$, we show there exists a particular positive integer value of $m$ that has minimum energy. This is then compared with the values observed in the experiments of Figs.~\ref{fig:voidexperiments} and \ref{fig:1periodexample}, using estimated values for all parametric constants, with some degree of success: the analytical model offers good agreement with the experimental values for $m$, and demonstrates that understanding the periodic formation of voids can be intimately linked to the balance of all contributing energies in tight geometric singularities. 

\smallskip
This simple extension of previous work on the single layered mode to multilayered configurations can be seen as a first step to the full multilayered model, including the complex geometric constraints arising from the layers being obliged to fit together. We believe that these highly nonlinear relations drive many of the multilayered patterns observed in true geological structures, and would be essential to any modelling programme.

\section{Method}
\label{sec:mperiodic}
In this section we set out the model described in this paper, and construct a total potential energy function for the system. An equilibrium equation is then derived using standard variational methods. 
\subsection{The model}
\label{subsec:modelling}
We consider an $N$ layered section of a larger chevron fold, as seen in Fig.~\ref{fig:chevronfold}, characterised by identical elastic layers of length $L$, uniform thickness $h$, elastic bending stiffness $EI$ and straight limbs at a constant angle $\varphi$ to the horizontal, connected by convex symmetric segments. The $N$ layered stack is arranged into an $m$-periodic solution, defined as a sequence of layers that repeats every $m^{th}$ layer with one void per periodic repeat; between voids, layers are constrained together to act as {\em parallel folds} (Budd et al. 2003).

\begin{figure}
\includegraphics[width=4in]{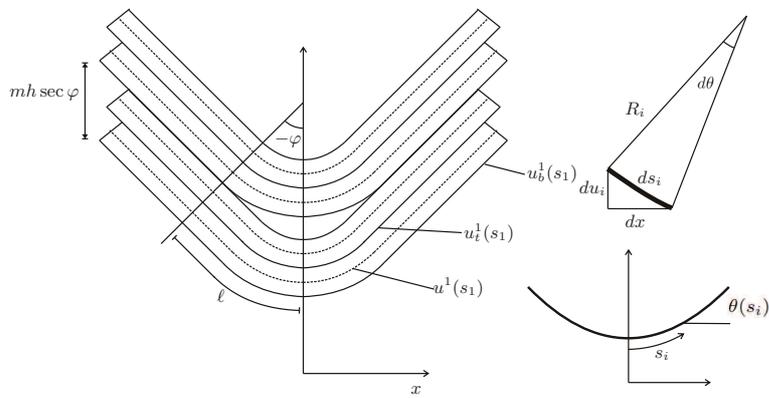}
\caption{Two 2-periodic packages, together with parameters and measurements used in the model $\theta(s_i)$.}
\label{fig:model}
\end{figure}

\smallskip
The setup and parameters of the model are summarised in Fig.~\ref{fig:model}. The centerline or neutral axis of the $i^{th}$ layer from the bottom of an $m$-periodic stack is defined by the vertical displacement $u^i(s_i)$, where $s_i\in [-L/2,L/2]$ is the arc length  of the $i^{th}$ layer, measured from the center of the fold.  Displacements of the top and bottom boundary of the layer are labelled $u^i_t(s_i)$ and $u^i_b(s_i)$, respectively. Intrinsic coordinates are characterised by $(s_i,\theta_i)$, where $\theta_i(s_i)$ is the angle the $i^{th}$ layer makes to the horizontal. It proves useful also to define the fixed horizontal coordinate $x$ within a global cartesian coordinate frame. As an aside, both representations have their uses: intrinsic coordinates give simpler equations, but for multilayered problems they carry the added complication of a different coordinate system for each layer. As seen later, it proves useful to be able to switch freely between coordinate frames. We also note that, since the assumption is made that the profiles are convex, for any layer there is a one-to-one-to-one relationship between $\theta_i$, $s_i$ and  $x$. The \emph{contact set} of an $m$-periodic structure is defined as the set $\Gamma_{} = \{s : u^{1}_b(s_1) = u^{m}_t(s_m) - mh\sec\varphi\}$, whereas its complement is the non-contact set $\Gamma^c$. Since profiles are assumed convex, the non-contact set is a single interval (Thm 2.2, Dodwell et al. 2012). The size of the void \emph{or} non-contact set is characterized by the free parameter $\ell = \ell_1 =\inf\{s>0:\Gamma\}$, with the corresponding horizontal measure $x_\ell$. Additionally we define the parameters $\ell_i = \inf\{s>0:u^i_b(s_i) = u_t^{i-1}(s_{i-1})\}$, for $i=2,\ldots, m$. Finally the shortening of a layer from a flat state is defined by the parameter $\Delta_i$. 

\smallskip
Throughout this work we make the assumption of~\emph{parallel folding}, such that layers in contact have a constant normal separation of distance $h$ between their neutral axes. Since by definition each layer within a single $m$-periodic package is in contact with its neighbours, its precise displacement may be calculated in terms of the outside layer $u_1$; for example, the position of layer $u_i$ can be determined by propagating layer $u_1$ a distance $ih$ upwards in its normal direction. This technique is used through the remainder of this paper, and means that the energy of an $m$-periodic package may be written in terms of single layer, $u_1$, simplifying the variational analysis which follows.

\smallskip
We now construct a total potential energy functional for this system, by considering the contributing energies in turn. Since the profile is assumed to be symmetric, we consider the energy over half the $m$-periodic stack, from $(0,\ell)$.

\subsubsection{Bending energy}

The total bending energy of a single $m$-periodic package is the sum of the bending energies of the individual layers. Classic bending theory (pp.28, Thompson \& Hunt 1973) gives the bending energy over a small segment of beam $ds_i$ as $dU_B = \frac 12EI\kappa(s_i)^2\;ds_i$, where $s_i$  and  $\kappa(s_i)$ are the arc length and curvature of the $i^{th}$ layer respectively. Integrating each $s_i$ over the curved section of each layer, the total bending energy in an $m$ periodic stack is given by,
\begin{equation}
U_B = \frac{EI}{2}\sum^{m}_{i=1}\int^{\ell_i}_{0}\kappa_i(s_{i})^2\;ds_i = \frac{EI}{2}\sum ^{m}_{i=1}\int^{\varphi}_{0}\kappa_i(\theta)\;d\theta.
\end{equation}
We now proceed to rewrite the bending energy, $U_B$, in terms of the outside layer $u_1$.  By noting that $R_i = R_1 - (i-1)h$ and $\kappa_i = 1/R_i$ it follows that
\begin{equation}
\kappa_i = \frac{\kappa_1}{(1 - (i-1)h\kappa_1},
\end{equation}and therefore
\begin{equation}
U_B = \frac{EI}{2}\sum^m_{i=1}\int^{\varphi}_{0}\frac{\kappa_1}{(1 - (i-1)h\kappa_1)}\;d\theta = \frac{EI}{2h}\int^{\varphi}_{0}\underbrace{\sum^m_{i=1}\frac{h\kappa_1}{1 - h(i-1)\kappa_1}}_{:=f(\kappa_1)}\;d\theta.
\end{equation}
 We then  consider $f(\kappa_1)$ to be a Riemann sum (Thomas 2006), and approximate it by the corresponding integral; i.e. setting $\xi_i = h(i-1)$, we approximate
\[ 
\sum^m_{i=1}\frac{h\kappa_1}{1 - \xi_i\kappa_1}
\quad\mbox{by}\quad
\int_{\xi_1}^{\xi_{m+1}} \frac{\kappa_1}{1 - \xi\kappa_1}\, d\xi 
= - \log (1-\xi\kappa_1)\Big|_{\xi=\xi_1}^{\xi=\xi_{m+1}},
\]
and therefore,
\begin{equation}
U_B = -\frac{EI}{2h}\int^{\varphi}_{0} \log(1 - H\kappa_1)\;d\theta = -\frac{EI}{2h}\int^{\ell}_{0} \log(1 - H\kappa_1)\kappa_1\;ds_{1},
\label{eqn:bending}
\end{equation}
where $H = mh$. The buckling energy is thus a strongly nonlinear function of the curvature and hence displacement $u_1$ of the reference layer. The important observation is that the bending energy becomes infinite as $\kappa_1$ approaches $1/H$, with logarithmic blowup. As a result, particularly sharp profiles, or large values of $H$,  carry a severe energy penalty.

\subsubsection{Work done against overburden pressure}
\begin{figure}
\centering
\includegraphics[width = 5in]{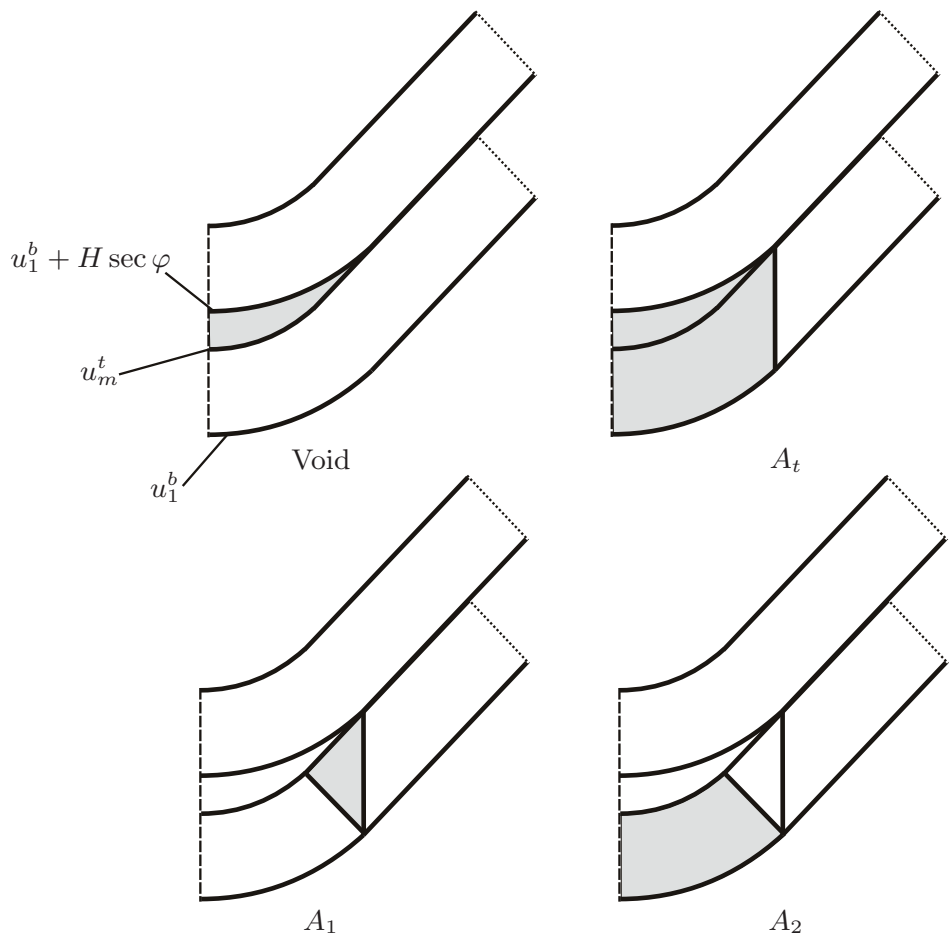}
\caption{Areas constructed to calculate the size of the void between two $m$-periodic packages, given by $A_t - A_1 - A_2$.}
\label{fig:voidgeometry}
\end{figure}

\begin{figure}
\centering
\includegraphics[width=3in]{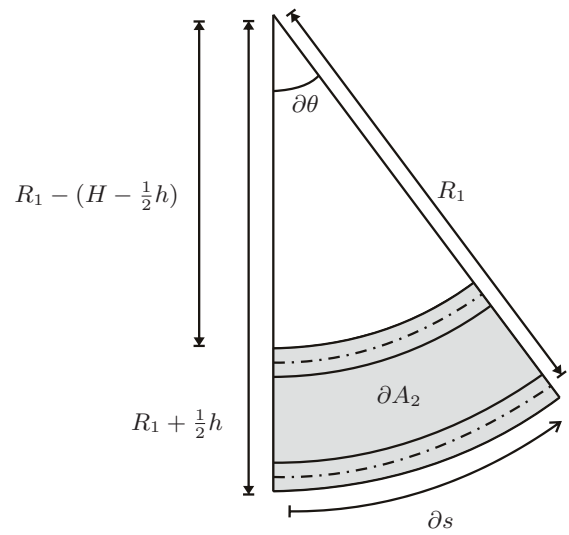}
\caption{Diagrams shows the construction of a small segment, $\partial A_2$}
\label{fig:segmentvoid}
\end{figure}

In experiments with paper, and their geological counterparts, the formation voids clearly involves a subtle interplay between release in overburden pressure in these areas, together with some compensatory increase in internal strain energy elsewhere. Rather than attempt to model this complex situation directly, we simply assume here that the creation of all voids does work against a constant pressure $q$ per unit length, regardless of position in the sample. For an $m$-periodic solution, by definition, a single void forms between the $1$st and $m$th layer, and to determine this area it is convenient to construct the series of complementary areas shown in Fig.~\ref{fig:voidgeometry}. Taking these in turn, area $A_t$ is given by
\begin{equation}
A_t = \int^{x_\ell}_{0} u^b_1 + H\sec\varphi - u^b_1\;dx = H\sec\varphi\int^\ell_{0}\cos \theta\;ds_{1}.
\end{equation}
Following a simple geometric construction, the area of triangle $A_1$ is equal to $\frac 12kH^2$. Finally we construct $A_2$, the area enclosed between the bottom of $u_1$ and the top of $u_m$. The area of a small segment $dA_2$ can be calculated using the geometry of a circular arc, Fig.~\ref{fig:segmentvoid}, given by
\begin{multline}
d A_2 =\frac 12\left((R_{1}(s) + \frac 12h)^2
 - (R_{1}(s) + \frac 12 h - H)^2\right)\;d\theta \\
 =\left(R_1(s)H +\frac{1}{2}hH - \frac{1}{2}H^2 \right)\;d\theta.
\end{multline}
Integrating over the curved section from $0$ to $\varphi$, we obtain
\begin{equation}
A_2 = \int^{\varphi}_{0} R_1(s)H + \frac{1}{2}(hH - H^2) \;d\theta= \int^\ell_{0}H+ \frac{1}{2}(hH - H^2)\kappa_1\;ds_{1}.
\end{equation}
The total size of the void is given by $A_t - A_1-A_2$, and hence the total work done against overburden pressure is
\begin{equation}
U_V = q\left[ H\sec\varphi\int^\ell_{0}\cos \theta\;ds_{1} - \frac 12kH^2 - \int^\ell_{0}H+ \frac{1}{2}(hH - H^2)\kappa_1\;ds_{1}\right].
\end{equation}

\subsubsection{Work done by axial compressive loads}
For the system to remain conservative, the loads at the end of each $m$-periodic package must remain both constant and at a fixed angle $\varphi$ to the horizontal. This means that work done by such a load can be measured from any position along its line of action, the difference between two alternative positions being a constant. We choose here to measure the corresponding deflection of a load from the position where its line of action meets the centre line $x=0$ of Fig.~\ref{fig:model}. As long as  $\varphi$ remains constant (unlike the model in Dodwell et al. 2012c), this has some analytical advantages (Dodwell et al. 2012). 
\smallskip

The total work done by these loads, per periodic package, thus can be written as $U_P = P\sum_{i=1}^m \Delta_i$, where
\begin{equation}
\Delta_i = \sqrt{1 + k^2}\int_0^{\ell_i}\cos\theta\; ds_i - \ell_i = \int_0^{\varphi}\left(\sqrt{1 + k^2}\cos\theta - 1\right)R_i\;d\theta.
\end{equation}
It follows that
\begin{equation}
U_P = P\sum_{i=1}^m \int_0^{\varphi}\left(\sqrt{1 + k^2}\cos\theta - 1\right)(R_1 - (i-1)h)\;d\theta.
\end{equation}
Given the sum of $n$ natural numbers is $\sum^n_{i=1} i= \frac 12n(n+1)$, it follows
\begin{align}
U_P &= P \int_0^{\varphi}\left(\sqrt{1 + k^2}\cos\theta - 1\right)\sum_{i=1}^m R_i\;d\theta \nonumber\\ 
&= P \int_0^{\varphi}\left(\sqrt{1 + k^2}\cos\theta - 1\right)\left(mR_1 + \frac 12hm(1-m)\right)\;d\theta.
\end{align}
Changing variables finally yields,
\begin{equation}
U_P = mP \int_0^{\ell_1}\left(\sqrt{1 + k^2}\cos\theta - 1\right)\left(1 + \frac 12h(1-m)\kappa_1\right)\;ds_1.
\end{equation}

\subsection{Total potential energy and Euler-Lagrange equation}
\label{subsec:TPEandEL}
The total potential energy of a stack of $N$ layers in an $m$-periodic structure is the sum of the contributing terms derived above, multiplied by $N/m$, the number of $m$-periodic packages in the $N$ layered stack. We write the potential in terms of the variable angle $\theta(s_{1})$, and note the relationship $\kappa_{1} = \theta^\prime(s_{1})$, where the prime denotes differentiation with respect to arc length $s_{1}$. As a point of notation we now drop the subscript 1, so the arc length of the bottom layer and its displacement are henceforth referred to as  $s$ and $u$ respectively.  The total potential energy is given by the functional 
\begin{eqnarray}\label{eqn:potential}
V(\theta) & = & \int_0^\ell\mathcal L(\theta,\theta^\prime)\;ds =  -\frac{EIN}{2H}\int^\ell_{0} \log(1 -H\theta^\prime)\theta^\prime\;ds \nonumber\\
& + & \frac{qN}{m}\left[ H\sec\varphi\int^\ell_{0}\cos \theta\;ds - \frac 12kH^2 - \int^\ell_{0}H+ \frac{1}{2}(hH - H^2)\kappa_1\;ds\right]\nonumber\\
  & + & PN \int_0^{\ell}\left(\sqrt{1 + k^2}\cos\theta - 1\right)\left(1 + \frac 12h(1-m)\theta'\right)\;ds.
\end{eqnarray}
Minimising solutions can be found by computing the derivative of $V$ with respect to small perturbations $\delta \theta$ of the function $\theta$, which gives
\begin{equation}
\delta V = \int_0^\ell \left(\frac{\partial \mathcal L}{\partial \theta^\prime}\delta \theta^\prime + \frac{\partial \mathcal L}{\partial \theta}\delta \theta \right)\;ds.
\end{equation}
Integrating once by parts yields the expression
\begin{equation}
\delta V = \left[\frac{\partial \mathcal L}{\partial \theta^\prime}\delta \theta\right]_0^\ell + \int_0^\ell \left(- \frac{d}{ds}\left(\frac{\partial \mathcal L}{\partial \theta^\prime}\right)+ \frac{\partial \mathcal L}{\partial \theta} \right)\delta \theta\;ds.
\end{equation}
Stationary solutions satisfy $\delta V = 0$ for all admissible perturbations $\delta \theta$, i.e. perturbations which satisfy the boundary condtions~\eqref{eqn:boundary}, such that
\begin{equation}
\delta\theta(0) = \delta\theta^\prime(\ell) = 0 \mbox{ and } \delta\theta(\ell) = \varphi.
\end{equation}
Stationarity is therefore equivalent $\theta$ solving the Euler-Lagrange equation
\begin{equation}
-\frac{d}{ds}\left(\frac{\partial\mathcal L}{d\theta^\prime}\right) + \frac{\partial \mathcal L}{\partial\theta} = 0,
\end{equation}
over the non-contact set $\Gamma = (0,\ell)$. We note that the boundary terms in the square brackets are zero from applying the free boundary conditions
\begin{equation}
\theta(\ell) = \tan^{-1}k = \varphi ,\quad \theta^\prime(\ell) = 0 \quad \text{and} \quad \theta(0) = 0.
\label{eqn:boundary}
\end{equation}
Appropriate differentiation of $\mathcal L$ yields the Euler-Lagrange equation
\begin{equation}
-\frac{1}{2}EI\frac{\theta''(2 - H\theta')}{(1 - H\theta')^2} - \left(qh\sec\varphi + P\sqrt{1 + k^2}\right)\sin\theta = 0.
\end{equation}
By noting that $2 - H\theta' > 1$ from the finiteness of~\eqref{eqn:bending}, it follows that
\begin{equation}
 N(s,\theta) = \theta'' + \lambda\sin\theta\frac{(1 - H\theta')^2}{2 - H\theta'} = 0,
\label{eqn:EL}
\end{equation}
where 
\begin{equation}
\lambda = \frac{2(qh\sec\varphi + P\sqrt{1 + k^2})}{EI}> 0.
\label{eq:lambda}
\end{equation}
This nonlinear second-order differential equation in $\theta$ can be physically interpreted as a force balance (Sec. 3, Dodwell et al. 2012), where $\theta^{\prime\prime}$ is the shear force acting in the outside layer normal to its neutral axis $u_1$.  By noting $\theta(s) = u_x$, \eqref{eqn:EL} can be written as a third-order differential equation in terms of deflection $u$, as for other examples of buckling and bending of a layer (Thompson \& Hunt 1973)
\smallskip 

In~\cite{Dodwell2011b}, constant curvature assumptions were used to obtain approximate solutions, but at the expense of contravening boundary conditions at points of layer separation $s = \pm \ell$. The fourth correct free boundary conditions obtained from~\eqref{eqn:EL} is,
\begin{equation}
 \theta^{\prime\prime}(\ell) = -\lambda\sin\varphi.
\end{equation}
This represents a discontinuous shear force, rather than a moment as assumed in Dodwell et al. (2012b), which reacts against the overburden pressure $q$ at the point lift-off. The same effective point-load reactive force can be seen in other similar examples involving lift-off, for example upheaval buckling of pipelines (Hunt \& Beardmore 1997) and single layered void formation (Dodwell et al. 2012).

\subsection{Numerical scheme for nonlinear boundary value problem}
\label{subsec:numerics}
\begin{figure}
\centering
\includegraphics[height = 1.9in]{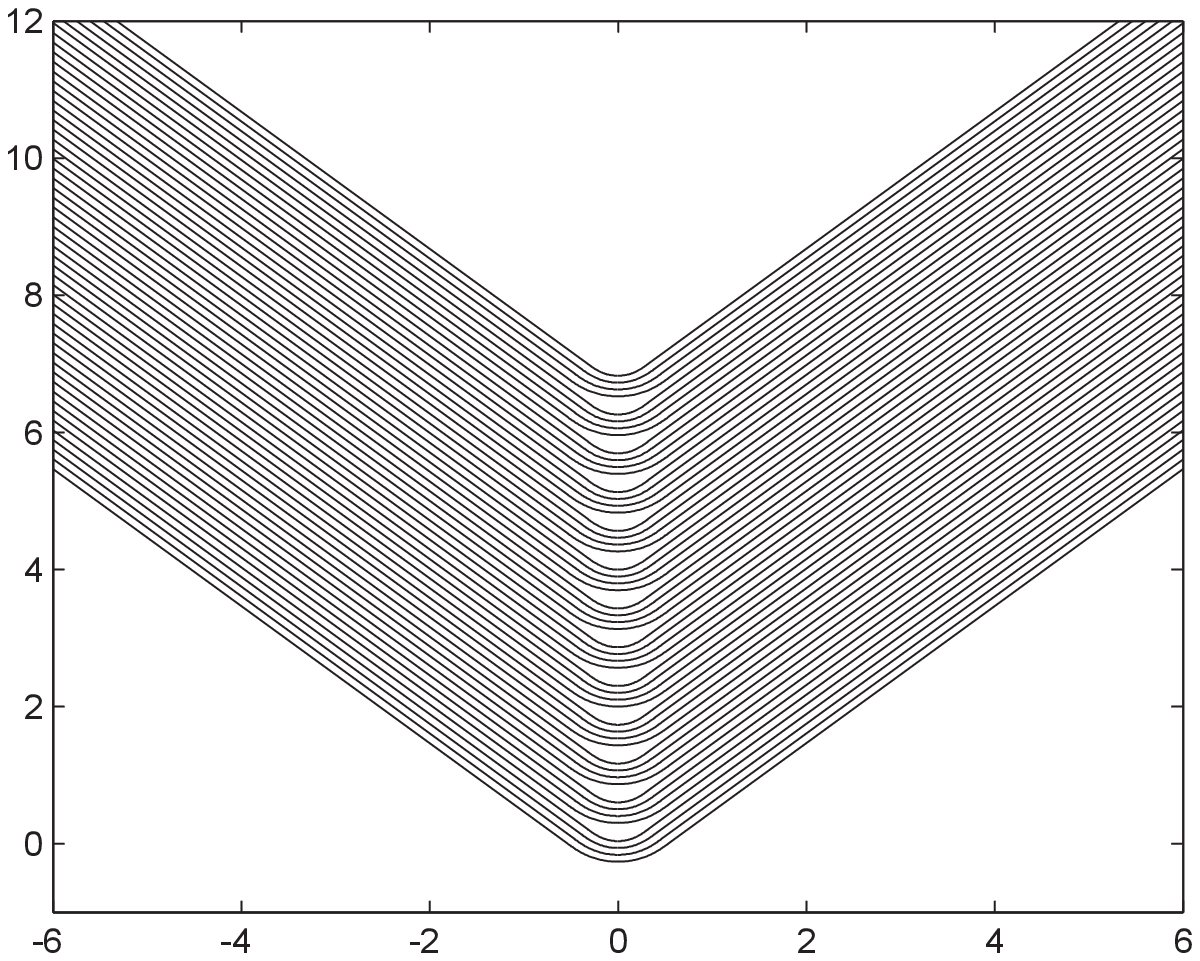}
\includegraphics[height = 1.9in]{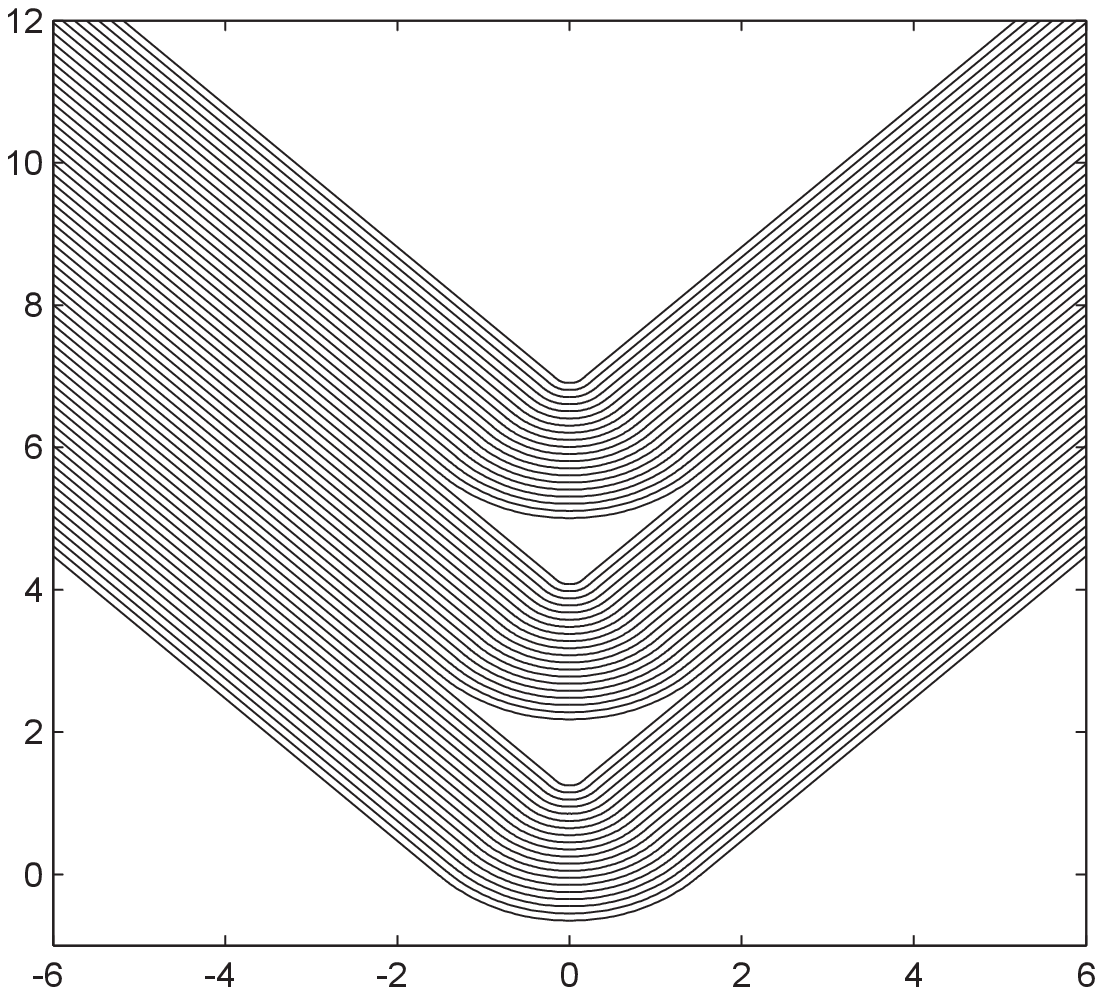}
\caption{Two examples of solution profiles for fixed values of $k = q = P = E = 1$, $h = 0.1$.  Left: $m = 3$ and $N=39$. Right: $m = 20$ and $N = 60$.} 
\label{fig:profiles}
\end{figure}

Numerical solutions of the nonlinear second-order boundary value problem are found by combining the collocation method (Kierzenka and Shampine 2008) and a homotopy perturbation technique. The homotopy process is used to introduce the nonlinear terms gradually  from a known linear solution. It ensures numerically stability, and is particularly important in this case since $\varphi$, the angle of the limbs, is typically large  (e.g. $\varphi = \frac{1}{4}\pi$). The system is first rescaled such that $s = \ell\hat{s}$, so~\eqref{eqn:EL} becomes
\begin{equation}
\theta^{\prime\prime}(\hat{s}) =   -\ell\lambda\sin\theta \frac{(\ell - H\theta^\prime)^2}{2\ell - H\theta^\prime},
\end{equation}
with primes now denoting differentiation with respect to $\hat{s}$. The problem is defined over the fixed domain, $\hat{s} \in [0,1]$,  and together with three boundary conditions
\begin{equation}
\theta(0) = 0,\quad \theta(1) =  \varphi \quad \mbox{and} \quad \theta^\prime(1) = 0,
\end{equation}
allowing $\ell$ to be free parameter found as part of the solution. The {\em homotopy} operator is defined as
\begin{equation}
\mathcal H(\hat{s},\theta_i,\zeta_i) = (1 - \zeta_i) L(\hat{s},\theta_i) + \zeta_i  N(\hat{s},\theta_i)
\end{equation}
where $\zeta_i$ is a finite sequence from $0 \rightarrow 1$, $N(\hat{s},\theta_i)$ is the nonlinear operator defined by \eqref{eqn:EL}, and $L$ is its linearisation. We note that if both $\theta$ and $\theta'$ remain sufficiently small, $N$ reduces to
\begin{equation}
 L (\hat{s},\theta) = \theta'' -\lambda\ell^2\theta= 0. 
\end{equation}
For the first step of the homotopy process, $i=1$, we solve the linearised problem so that $\zeta_1 = 0$ and  $$\mathcal H(\hat{s},\theta_1,\zeta_1) =  L(\hat{s},\theta_1).$$ Paired with the boundary conditions~\eqref{eqn:boundary}, this has the solution
\begin{equation}
\theta_1(\hat{s}) = \frac{\varphi\ell^2}{\sin \sqrt{\lambda}\ell}\sin \sqrt{\lambda}\hat s. \quad \mbox{and} \quad \ell = \frac{1}{2}\pi \lambda^{-1/2}.
\label{eqn:linearsolution}
\end{equation}
For sequential steps, $i > 1$, $\mathcal H(\hat{s},\theta_i,\zeta_i)$ is solved using collocation. Here we used \textsc{MATLAB}'s in-built function \texttt{bvp4c},  providing $\theta_{i-1}$ and 
\begin{equation}
\ell_i = \ell_{i-1} + \left(\zeta_i - \zeta_{i-1}\right)\left(\frac{\ell_{i-1} - \ell_{i-2}}{\zeta_{i-1} - \zeta_{i-2}}\right)
\end{equation}
as initial guesses. Insuring the increments of $\zeta_i$ are sufficiently small, solutions to the full nonlinear problem are found. Solution profiles for two sets of parameters values are shown in Fig.~\ref{fig:profiles}. Numerical evidence suggests that there is a unique convex solution to (\ref{eqn:EL}); however a mathematical proof remains an open question.

\subsection{Scaling of $\ell$ with respect to other parameters}
Once the solution $\theta$ is found using the method described in the previous section, the variation of the solutions with various physical parameters can be investigate using continuation methods.  We now consider how the size of the void varies with variations in phyiscal parameters. Firstly we consider the scaling shown by the linear solution, $L = 0$. For this case the size of the void $\ell$ is fixed by imposing the boundary condition $\theta'(\ell) = 0$ (i.e. zero bending moment at the point of delamination) so that
\begin{equation}
\quad \ell = \frac{1}{2}\pi \lambda^{-1/2},
\label{eqn:ellscaling}
\end{equation}
where $\lambda$ is given by \eqref{eq:lambda}. The scaling matches physical intuition, in that as the overburden pressure $q$ or the axial load is $P$ increased, the size of the void $\ell$ decreases. Likewise, void size increases with increasing bending stiffness $EI$. 

\section{Application}

\subsection{Minimum energy $m$-periodic solutions}\label{sec:mmin}

The numerical method outlined in the previous section allows one to calculate a range of solutions over different values of $m$,  for each of which the potential energy (\ref{eqn:potential}) can be determined numerically. We can then answer two linked questions: first, for a given set of  parameters, what value of $m$ gives the minimum energy configuration? Secondly: how does this value, which we will call $m^*$, alter with respect to overburden pressure $q$ and axial load $P$? 

\smallskip
For this we need not only estimates of the material parameters used in the model, but also the axial and overburden loads for the configurations seen in Figs. \ref{fig:voidexperiments} and \ref{fig:1periodexample}. Applied axial loads were registered on load cells at the time, but overburden pressures were not recorded in these particular tests and provide more of a problem. Where they have been documented, as for example in the kink-band tests of Wadee et al. (2004), they seem to show the following property: after the initial instability, {\em total} axial and transverse loads are approximately equal, and increase at approximately the same rate. In the absence of any further information, we shall adopt this criterion as a reasonable best guess: knowing the axial load, we divide by the paper area to obtain an approximation of the overburden pressure. 

\smallskip
We therefore adopt the simple relationship $q = P/A$ between overburden pressure and axial load, where $A$ is the area of the sides supported by the foam foundation. If we revert attention to the experiments shown in Fig.~\ref{fig:voidexperiments} (Experiment 1) and Fig.~\ref{fig:1periodexample} (Experiment 2), we see that  Experiment 2 is pushed much further into the post-buckling range than Experiment 1; consequently the axial load, and therefore the constraining overburden stress, is an order of magnitude higher, at approximately $30$ kN compared with $< 1$ kN. 

\begin{figure}
\centering
\includegraphics[width = 5in]{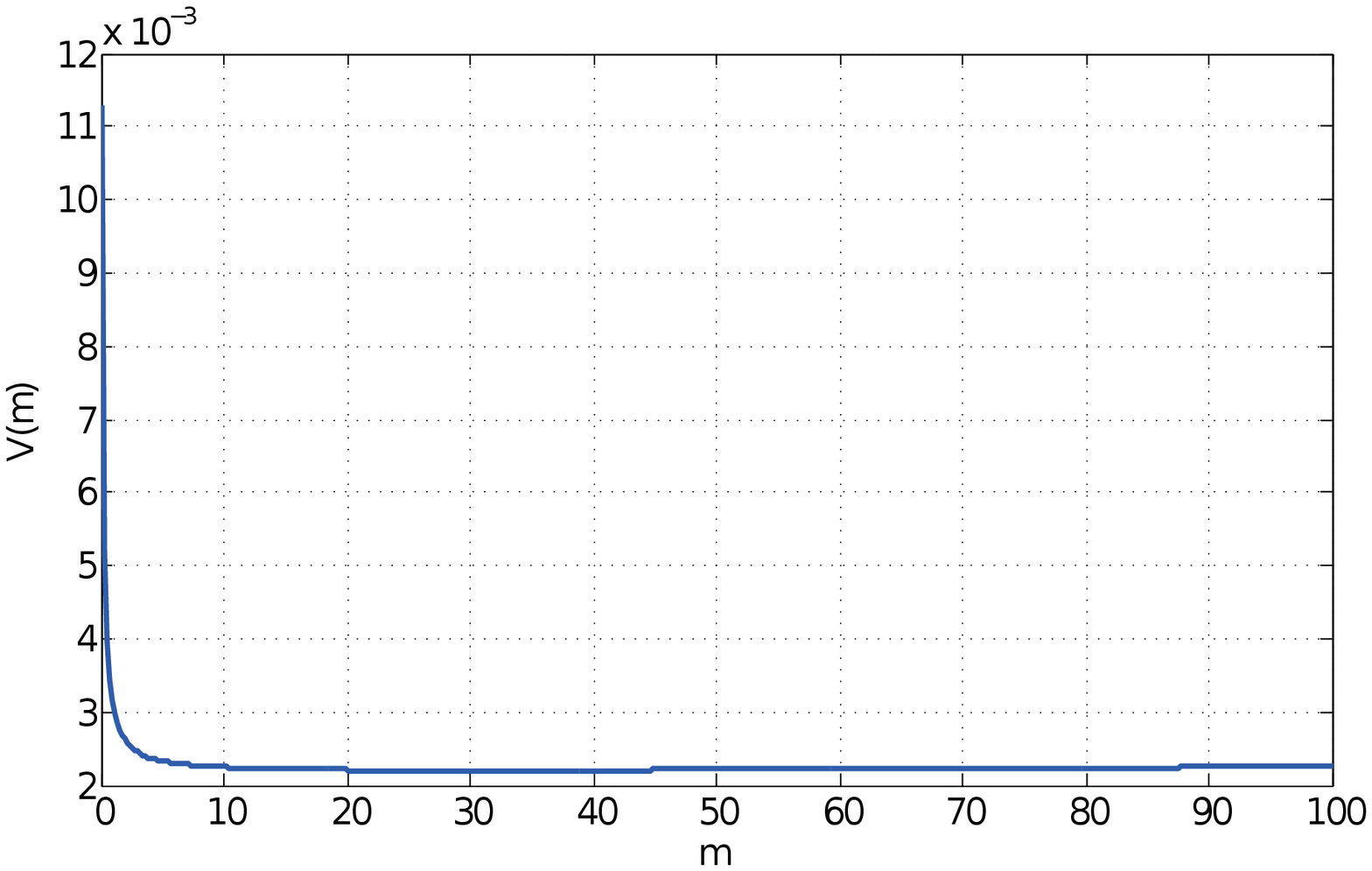}\quad
\includegraphics[width = 5in]{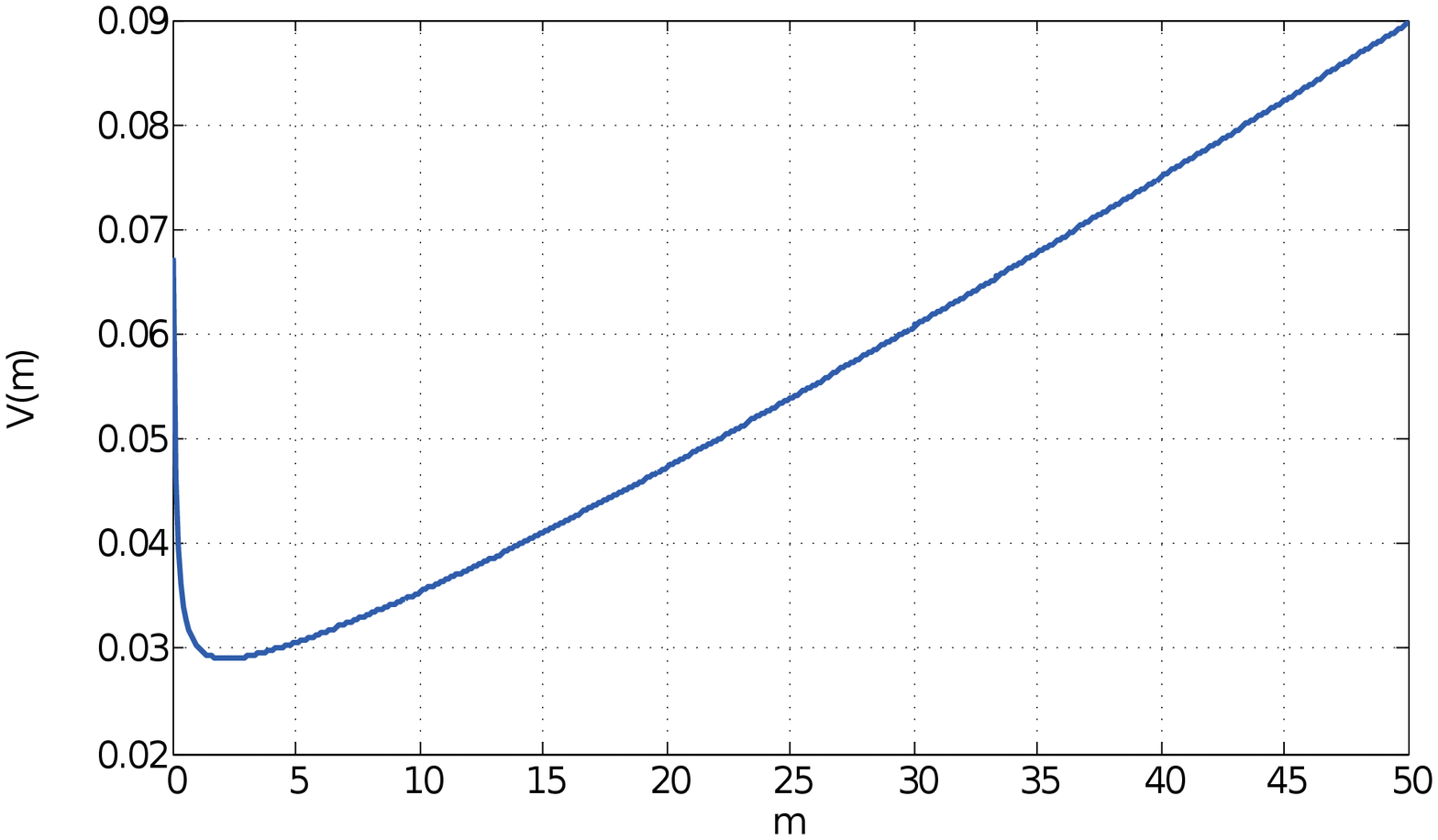}
\caption{Variation of potential energy $V$ with corresponding void number $m$, for the two experiments of Figs.~\ref{fig:voidexperiments} and Fig.~\ref{fig:1periodexample}, using the estimated parameter values of Table~\ref{tab:parameters2}.} 
\label{fig:compexpts}
\end{figure}

\smallskip
We now investigate the change in theoretical value of $m^*$ for each experiment, with respect to changes in axial load and hence overburden pressure. Do the ranges of values of loads recorded correlate to the observed values of approximately $m^* = 125$ for Experiment 1, and $m^* =1$ for Experiment 2? Table \ref{tab:parameters2} shows parameter value estimates for each experiment, from measurements taken during the course of each. Fig~\ref{fig:compexpts} shows the result of plotting load $P$ against the corresponding value of $m$ for each circumstance. 

We see that in each case the model gives a reasonable outcome when compared with experiment. For Experiment 1, the energy reaches a shallow minimum at about $m = 40$; this seems reasonable, given the shallowness of the minimum and  disturbances introduced by blackening the edge of one sheet of paper in every 25. For Experiment 2, the lowest integer value of $m$ appears to be $m^* = 2$;  although the actual minimum occurs at about $m = 2.3$, this is a non-integer value and is thus denied to the model. It is significant that $V \rightarrow \infty$ as $m \rightarrow 0$, so the absence of voids is not an option. 

\section{Conclusions}

This paper highlights the importance to the multilayered systems that appear in many stratified geological situations, of geometric constraints that come from layers being obliged to fit together or nearly so. The resulting deformations come about as a subtle interplay of these geometric constraints and the mechanical properties of the layers themselves.  This contribution extends a previous model for $m$-periodic structures (Dodwell et al. 2012), by satisfying the correct boundary conditions at points of layer separation and solving the resulting nonlinear differential equation numerically. The simple energy-based argument  describes the periodic formation of voids in the hinge region of chevron folds, where the contributing potential energies are balanced to give a $m$ periodic structure with minimum energy. Given that parameter values, particularly for the overburden pressure, are difficult to estimate, the method  provides reasonable agreement with two simple experiments on constrained sheets of paper.

\smallskip
The model is of course highly simplified, and many modifications and generalizations can be envisaged. An important assumption is the pure elasticity of the layers, where there is good reason to consider other material properties; sharp corners in a bending problem will often imply the formation of plastic hinges, for example Dodwell et al. 2012c. The advantage of such a simplified study, however, is that the appearance and size of voids are precisely defined, and the assumptions of elasticity allow precise analysis.  

\smallskip
The  inclusion of friction in future models is clearly important, as during the bending and voiding process layers will want to slip over each other. It will be vital to embrace not only the complex topological effects of layers fitting together, but slipping relative to each other (Hobbs et al . 1976). The role of friction will be crucial to multilayered modelling in the future, and will be especially important in the formation of voids. However, as the frictional forces can switch  direction (Budd et al. 2003), and the normal contact force can vary as the voids develop (Hunt et al. 2013), analysis is likely to become considerably more complicated and energy-based approaches perhaps of less immediate use.

\begin{acknowledgements}
The authors would like to thank both Mark Peletier and Bruce Hobbs for various discussions and inputs throughout this work, as well as Ahmer Wadee who during his time at Bath carried out Experiment 1 under the EPSRC project GR/L17177/01.
\end{acknowledgements}


\bibliographystyle{spbasic}

\newpage

\begin{table}
\begin{center}
\begin{tabular}{|c|c|c|c|}\hline\centering \textbf{Parameters} & \textbf{Experiment 1} & \textbf{Experiment 2} & \textbf{Source} \\ \hline
 $E$ & 5 kNmm$^{-2}$& 5 kNmm$^{-2}$& Wadee et. al 2004
\\
 $h$ & 0.08 mm& 0.08 mm&Wadee et. al 2004\\
 $b$ & 105 mm& 148 mm& Dimension of A$6$ and A$5$ paper\\
 $I$ & $4.48\times10^{-3}$ mm$^{-4}$ & $6.31\times10^{-3}$ mm$^{-4}$ & $h^3b/12$ \\
$N$ & 500 & 750 & - \\
$A_1$ & 4,200\;mm$^2$& 8,880\;mm$^2$& $b$ multiplied by $hN$ \\
$A_2$ & 15,540\;mm$^2$& 31,080\;mm$^2$& Dimension of A$6$ and A$5$ paper \\$k$ & $ 0.25$ & $1$& From Fig.~\ref{fig:voidexperiments} and Fig.~\ref{fig:1periodexample} \\ 
$P$ & $0.5$ & $30$ & Recorded during experiment\\ \hline
\end{tabular}
\end{center}
\caption{Parameters values for each of the experiments.}
\label{tab:parameters2}
\end{table}

\end{document}